\documentclass[10pt,a4paper,twocolumn,english,prb,aps,floatfix,showpacs,groupedaddress,superscriptaddress]{revtex4-2}

\usepackage{graphicx}
\usepackage{epsfig}
\usepackage[english]{babel}
\usepackage{amsmath}
\usepackage{amssymb}
\usepackage{amsfonts}
\usepackage{dcolumn}
\usepackage{bm}
\usepackage{bbm}
\usepackage{nicefrac}
\usepackage{color,array}
\usepackage{colortbl}
\usepackage{hyperref}
\usepackage[caption=false,position=top]{subfig}
\usepackage{dsfont}

\def\Bext{B_\mathrm{ext}}

\def\ez{\bm{e}_z}
\def\lambg{\lambda_\mathrm{g}}
\def\lambt{\lambda_\mathrm{t}}
\def\tausg{\tau_\mathrm{s,g}}
\def\taust{\tau_\mathrm{s,t}}
\def\tauseff{\tau_\mathrm{s}^*}

\def\S{\bm{S}}
\def\T{\bm{T}}

\def\wn{\omega_\mathrm{n,g}}
\def\wm{\omega_\mathrm{m}}
\def\wt{\omega_\mathrm{n,t}}

\def\TR{T_\mathrm{R}}
\def\Tm{T_\mathrm{m}}

\def\fm{f_\mathrm{m}}
\def\geff{\mathrm{eff,g}}
\def\teff{\mathrm{eff,t}}
\def\e{\mathrm e}
\renewcommand{\d}{\mathrm d}

\begin{document}

\title{Spin inertia and polarization recovery in quantum dots:\\Role of pumping strength and resonant spin amplification}

\author{Philipp Schering}
\email{philipp.schering@tu-dortmund.de}
\author{G\"otz S. Uhrig}%
 \email{goetz.uhrig@tu-dortmund.de}
\affiliation{%
	Lehrstuhl f\"ur Theoretische Physik I, Technische Universit\"at Dortmund, Otto-Hahn-Straße 4, 44221 Dortmund, Germany
}%

\author{Dmitry S. Smirnov}
\email{smirnov@mail.ioffe.ru}
\affiliation{
Ioffe Institute, 194021 St.~Petersburg, Russia
}%

\date{\today}

\begin{abstract}
Spin inertia measurements are a novel experimental tool to study long-time spin relaxation
processes in semiconductor nanostructures. We develop a theory of the spin inertia effect for resident electrons and holes localized in quantum dots. 
We consider the spin orientation by short optical pulses with arbitrary pulse area and detuning from the trion resonance.
The interaction with an external longitudinal magnetic field and the hyperfine interaction with the nuclear spin bath is considered in both the ground and excited (trion) states of the quantum dots.
We analyze how the spin inertia signal depends on the magnetic field (polarization recovery) and on the modulation frequency of the helicity of the pump pulses as well as on their power and detuning. 
In particular, we elaborate how approaching the saturation limit of the spin polarization influences the measurements.
The quantitative description of spin inertia measurements will enable 
the determination of the parameters of spin dynamics such as the spin relaxation times in the ground and excited states and the parameters of the hyperfine interaction.
Finally, we predict the emergence of resonant spin amplification due to the transverse components of the nuclear spin fluctuations, which manifests itself as oscillations of the spin polarization as a function of the longitudinal magnetic field.%
\end{abstract}


\maketitle

\section{Introduction}
\label{sec:introduction}

The era of spin physics in semiconductors~\cite{dyakonov_book} has triggered the 
introduction of a number of new experimental techniques. 
The most fundamental one is the measurement of the Hanle effect~\cite{OptOr}, which allows 
one to determine the spin lifetime from the dependence of the degree of circular polarization 
of the photoluminescence on the transverse magnetic field. 
A more powerful technique based on pump-probe experiments appeared 
later \cite{PhysRevLett.55.1128,Zheludev1994823}, which rendered it possible to monitor the spin dynamics with picosecond time resolution. 
This technique has been extended to measure the long-time spin dynamics~\cite{Extended_pp} and merged with Hanle measurements leading to so-called resonant spin 
amplification (RSA)~\cite{Kikkawa98,yugov12}. 
In the past few years more modern methods have appeared aiming mainly at the measurement of 
slow spin dynamics on the time scale of microseconds. 
Examples are the spin noise spectroscopy~\cite{zapas13b} studying the spin fluctuations in 
thermal equilibrium and the spin inertia measurement~\cite{heisterkamp15}.
The latter is the subject of the present article.

Although the first spin inertia measurements were based on a discrete 
pump-probe technique~\cite{heisterkamp15}, 
the essence of the method can be understood considering 
continuous optical spin orientation~\cite{Ignatiev05,Fras2011}. 
The spin polarization induced by the application of circularly polarized
light pulses is measured. The spin inertia manifests itself upon modulation 
of the helicity of the circularly polarized light with a finite frequency $\wm$. 
When this frequency increases and becomes larger than the inverse spin relaxation time
the spin polarization decreases and eventually vanishes for very large $\wm$. 
This led to calling the effect \emph{spin inertia} because it can be seen as an inertia
of the spin polarization which prevents it from following an external switching of helicity
arbitrarily quickly.
The measurement of the dependence of the spin polarization on the modulation
frequency $\wm$ allows one to determine the spin relaxation time of the resident charge carriers.

Experimentally this method is useful to measure long spin relaxation times, for instance 
for localized electrons and holes in a longitudinal magnetic field~\cite{zhukov18}. 
Therefore, a natural object for this kind of study is an ensemble of localized charge carriers in quantum dots (QDs) or localized charge carriers bound to impurities in bulk semiconductors. 
The slow spin dynamics in such systems is mainly driven by the hyperfine interaction with the 
nuclear spins of the host lattice~\cite{book_Glazov}. 
The simple phenomenological model used for the description of the first 
experiments~\cite{heisterkamp15} was later extended to account for non-Markovian spin 
dynamics typical for localized electrons~\cite{smirnov18}. 
Despite the generality of this model and the possibility to describe the spin inertia for any Green's function of the spin dynamics
the effects caused by the saturation of the spin polarization were not analyzed in detail.

Motivated by the recent experiments presented in Ref.~\cite{zhukov18}, we develop a theory of the spin inertia effect for electrons and holes localized in QDs and account for arbitrarily strong pumping of the system. 
In particular, our model is used to numerically simulate and analytically describe the spin inertia effect for pumping by $\pi$ pulses. 
We take into account the interplay of the hyperfine interaction with the
nuclear spin bath and the external magnetic field in the ground and excited (trion) states of the system. 
Our simulations demonstrate the decrease of the effective spin relaxation time with an increase of the pump power, 
which was observed in pump-probe experiments~\cite{zhu06,greilich2012ZnSe:F}, 
in measurements of the spin noise~\cite{dahba14,noise-trions}, and of the spin 
inertia~\cite{heisterkamp15,zhukov18}. 
We also study the dependence of the spin polarization on the external longitudinal magnetic field and on the detuning of the pump pulses from the trion resonance. 
Finally, we predict a regime where resonant spin amplification occurs, i.e., the spin inertia signal resonantly depends on whether or not the Larmor precession period
of the localized carrier spins is commensurate with the time between consecutive pump pulses \cite{yugov12}.
Such a Larmor precession takes place in spite of orienting the localized carrier spin along
the longitudinal external magnetic field (Faraday geometry) due to the transverse nuclear spin fluctuations.

The paper is organized as follows. 
In the next section, we present the model which we use for numerical simulations of the spin 
inertia effect and its analytical description in limiting cases. 
In Sec.~\ref{sec:results}, we first consider the limit of a strong longitudinal 
external magnetic field where the hyperfine interaction can be neglected.
Then, we study the dependence of the spin polarization on the magnetic field and also 
on the detuning of the pump pulse.
Finally, we predict and describe the emergence of RSA in Faraday geometry. 
Our findings are briefly summarized Sec.~\ref{sec:conclusions} and details of our calculations can be found in Appendix~\ref{sec:methods}.

\section{Model}
\label{sec:model}

We consider an ensemble of singly charged quantum dots in a pump-probe type experiment~\cite{zhukov18}. 
The QDs can be charged either with electrons ($n$ type) or with holes ($p$ type).
These two cases will be treated on equal footing, but the parameters of the spin dynamics 
are drastically different. 
The pump pulses resonantly excite singlet trion states
leading to the resident charge carrier spin orientation according to the optical selection rules~\cite{ivchenko05a}. 
The spin polarization can be probed by weak linearly polarized pulses using the 
spin Faraday effect~\cite{glazo12b,yugov09}.

In measurements of the spin inertia, the QDs are excited by a long train of $M \gg 1$ pump pulses
following one another with the repetition period $\TR$. 
Each pulse is circularly polarized and the helicity of the pulses is alternated 
with the frequency $\wm$. The spin inertia effect manifests itself in the dependence of the spin polarization degree on $\wm$. 
The spin polarization is probed by weak probe pulses with the same repetition period.
Experimentally, they arrive shortly before the next pump pulse~\cite{zhukov18}. 
Theoretically, we assume an infinitesimal negative delay $\tau_{\mathrm{d}}=-0$.
The Faraday rotation of the probe pulses yields the spin polarization along the 
axis of light propagation, which is also the direction of the structural growth.
It is referred to as the $z$ axis in the following. The spin inertia signal is defined by~\cite{heisterkamp15,smirnov18}
\begin{equation}
\label{eq:L_def}
L(\wm) = \frac{1}{M} \left| \sum_{k=1}^{M} S^z(k\TR + \tau_\mathrm{d}) 
\mathrm{e}^{i \wm (k \TR + \tau_\mathrm{d})} \right|.
\end{equation}
Qualitatively, this expression describes the amplitude of the Fourier component 
of the spin polarization $S^z(t)$ at the modulation frequency $\wm$.

The goal of this paper is to describe the spin dynamics and to calculate 
$L(\wm)$ for various system parameters. In order to address experimentally relevant conditions, 
we take into account the hyperfine interaction and the external longitudinal magnetic field. 
In contrast to what has been done in Ref.~\cite{smirnov18}, we do not impose any restrictions 
on the system parameters. In particular, we do not assume the pump pulses to be infinitely weak.

The spin dynamics of the resident charge carrier between consecutive pump pulses is described by
\begin{equation}
\label{eq:dSdt}
\frac{\d\S}{\d t} = \left(\bm\Omega_\mathrm{N,g}+\bm\Omega_\mathrm{L,g}\right)\times\S 
- \frac{\S}{\tausg} + \frac{J^z}{\tau_0} \ez,
\end{equation}
where $\bm\Omega_\mathrm{N,g}$ is the frequency of the spin precession in the random nuclear spin bath, called the Overhauser field,
and $\bm\Omega_\mathrm{L,g}= \Omega_\mathrm{L,g}\ez = g_\mathrm{g} \mu_\mathrm{B} \Bext \ez / \hbar$ 
is the Larmor frequency, with $g_\mathrm{g}$ the effective longitudinal $g$ factor of the ground state,
$\mu_\mathrm{B}$ the Bohr magneton, and $\Bext\ez$ the external longitudinal magnetic field (Faraday geometry).
Furthermore, the phenomenological term $-\S/\tausg$ describes the spin relaxation
unrelated to the hyperfine interaction with the nuclear spins in the QD. 
Possible mechanisms are spin-orbit and electron-phonon interactions.
The vector $\bm J$ is the trion pseudospin and $\tau_0$ is the trion lifetime related
to radiative decay. The unit vectors $\bm e_\alpha$ point along the 
$\alpha$ axes, with $\alpha=x,y,z$. 
The last term in Eq.~\eqref{eq:dSdt} describes the input of spin polarization due to the
radiative trion recombination~\cite{yugov09,zhukov07}. 
In contrast to Ref.~\cite{smirnov18}, we neglect the dynamics of the Overhauser field and assume it to be statically frozen~\cite{merku02}.

Note that in the presence of an external longitudinal magnetic field, the Zeeman effect will lead to a residual spin component $S^z_0 \ez = \tanh[g_\mathrm{g} \mu_\mathrm{B} \Bext / ( 2 k_\mathrm{B} T)]/2 \,\ez$, where $T$ is the temperature and $k_\mathrm{B}$ is the Boltzmann constant. This could be included in the equations of motion, but it is negligibly small under the experimental conditions. 
For instance, in a field of $\Bext = 200$~mT at temperature $4$~K \cite{zhukov18}, the thermal equilibrium spin polarization is only about $1\%$. Furthermore, the thermal spin polarization increases the spin polarization slightly for one helicity of the pump pulse but decreases it for the other one so that the total effect cancels in linear order in the measured signal \eqref{eq:L_def}. 
Only in second order, a minute effect can occur.

Similarly to Eq.~\eqref{eq:dSdt}, the trion pseudospin dynamics between the pump pulses is described by
\begin{equation}
\label{eq:dJdt}
\frac{\d\bm J}{\d t}=\left(\bm\Omega_\mathrm{N,t}+\bm\Omega_\mathrm{L,t}\right)\times\bm J - \frac{\bm J}{\taust} - \frac{\bm J}{\tau_0},
\end{equation}
where the subscript t refers to the parameters of the spin of the unpaired charge carrier 
in the trion. 
Importantly, for resident electrons the trion consists of the two electrons in a spin singlet state
and a hole with unpaired spin, while for resident holes it consists of
two holes in a spin singlet state and an electron with unpaired spin. 
Thus, effectively, the type of the charge carrier in the excited state is opposite to the type of the charge carrier in the ground state. 
We also note that the nonradiative trion recombination, which does not contribute to the spin polarization in the ground state, can be accounted for by renormalization of the relaxation time $\taust$. 
Combined with Eq.~\eqref{eq:dSdt}, only the $z$ component of the trion pseudospin $\bm J$ is transferred back to the ground state $\bm S$ during the trion lifetime $\tau_0$, which follows from the optical selection rules for the optical transitions between the bands $\Gamma_6$ and $\Gamma_8$ in GaAs-based semiconductors \cite{zhukov07,glazo12b}.

The Overhauser field caused by the hyperfine interaction with the nuclear spins is normally distributed so that its probability distribution has the form~\cite{schulten76,merku02,stane14b}
\begin{multline}
\label{eq:distribution}
\mathcal{F}(\bm\Omega_\mathrm{N,g}) = \frac{\lambda_\mathrm{g}^2}{(\sqrt{\pi} \omega_\mathrm{n,g})^3}\\ \times \exp\left(-\lambda_\mathrm{g}^2\frac{(\Omega_\mathrm{N,g}^x)^2+(\Omega_\mathrm{N,g}^y)^2}{\omega_\mathrm{n,g}^2} - \frac{(\Omega_\mathrm{N,g}^z)^2}{\omega_\mathrm{n,g}^2}\right),
\end{multline}
where $\omega_{\mathrm{n,g}}^2$ describes the variance and $\lambda_\mathrm{g}$ is
the anisotropy degree of the hyperfine interaction. 
In the case of electrons $\lambda_\mathrm{g}=1$ holds, while for heavy holes 
$\lambda_\mathrm{g}>1$. 
The trion spin precession is related to the same nuclear spin bath and we assume that it is given by
\begin{subequations}
	\begin{align}
	\Omega_\mathrm{N,t}^x &= \chi\frac{\lambda_\mathrm{g}}{\lambda_\mathrm{t}}\Omega_\mathrm{N,g}^x,
	\qquad 
	\Omega_\mathrm{N,t}^y = \chi\frac{\lambda_\mathrm{g}}{\lambda_\mathrm{t}}\Omega_\mathrm{N,g}^y,\\
	\Omega_\mathrm{N,t}^z &= \chi\Omega_\mathrm{N,g}^z,
	\end{align}
\end{subequations}
where $\chi=\omega_\mathrm{n,t}/\omega_\mathrm{n,g}$ describes the relative strength of the hyperfine interaction for the trion state and $\lambda_\mathrm{t}$ quantifies the anisotropy 
in the trion state. 
These relations are exact when the trion wave function is a product of identical wave functions of the three charge carriers, which holds true for small QDs~\cite{esser2000}.

Next we turn to the description of the pump pulses.
Experimentally, they have a duration of the order of 1~ps, which is typically much shorter than all the above time scales of the spin dynamics.
Then, the action of the pump pulse can be described by a relation between the spin components
before ($\bm S_\mathrm{b}$ and $\bm J_\mathrm{b}$) and after ($\bm S_\mathrm{a}$ and $\bm J_\mathrm{a}$) the pump pulse~\cite{yugov09,yugov12},
\begin{subequations}
	\begin{align}
	S^z_\mathrm{a} &= - \frac{\mathcal{P}}{4} (1 - Q^2) + \frac{1 + Q^2}{2} S^z_\mathrm{b}, \label{eq:pulse_Sz}\\	
	S^{x}_\mathrm{a} &= Q (S^{x}_\mathrm{b} \cos\Phi  + \mathcal{P} S^{y}_\mathrm{b} \sin\Phi ),\\
	S^{y}_\mathrm{a} &= Q (  S^{y}_\mathrm{b} \cos\Phi - \mathcal{P}  S^{x}_\mathrm{b} \sin\Phi),\\
	J^z_\mathrm{a} &= S^z_\mathrm{b} - S^z_\mathrm{a}, \label{eq:pulse_trion}\\
	J^{x}_\mathrm{a} &= 0, \qquad J^y_\mathrm{a} = 0.
	\end{align}
	\label{eq:pulse}%
\end{subequations}
Here, $0\le Q^2 \le 1$ is the probability \emph{not} to excite a trion by a pump pulse, 
$\Phi$ describes the spin rotation introduced by detuned pulses, and $\mathcal{P}=\pm1$ is the helicity of the pump pulse. 
Previous theoretical treatment assumed $Q$ close to 1~\cite{smirnov18}, corresponding to very weak pulses.
Here we are interested in arbitrarily strong pump pulses, in particular in $\pi$ pulses with $Q=0$.
In addition, we assume the trion state to be unoccupied just before the
subsequent pump pulse ($\bm J_\mathrm{b}=0$). This 
assumption is perfectly valid in the realistic case $\tau_0\ll \TR$.

In the particular case of zero modulation frequency ${\wm=0}$ it is not necessary to solve the equations for the spin dynamics for many pump pulses. 
Instead, one can use the fact that in this case the spin polarization is a periodic function 
with period $\TR$~\cite{glazo12b}. 
Provided $\bm S_\mathrm{b}$ is known, one can find $\bm S_\mathrm{a}$ and 
$\bm J_\mathrm{a}$ using Eq.~\eqref{eq:pulse}, 
and then solve Eqs.~\eqref{eq:dSdt} and~\eqref{eq:dJdt} on the time interval $t \in [0,\,\TR]$ for the given Overhauser field $\bm\Omega_\mathrm{N,g}$. 
The result has to coincide with $\bm S_\mathrm{b}$ because of the periodicity, which we exploit to calculate the steady-state value for $\bm S_\mathrm{b}$.
Finally, the spin inertia signal $L(\wm = 0)$ is proportional to the average of the steady-state value $S^z_\mathrm{b}$ over the distribution of the Overhauser field. 
Details of this calculation are described in Appendix~\ref{sec:methods_zero_fm}. 

Overall, we have checked that our results are in agreement with the results of Ref.~\cite{smirnov18} for weak pump pulses.
Slight deviations are found only when the assumptions used in Ref.~\cite{smirnov18} are not completely valid.

\section{Results}
\label{sec:results}

In this section, we analyze the spin inertia signal for various parameters and in particular for different pump powers. 
First, we focus on the case of a strong longitudinal magnetic field and resonant pulses. 
Then, we turn to the dependence of the spin polarization on the longitudinal magnetic field 
and on detuning. At the end of this section, we predict and describe the 
emergence of RSA in Faraday geometry.

\begin{table}[b]
	\caption{\label{tab:parameters} Choice of system parameters used for $n$- and $p$-type QDs, based on the measurements of Ref.~\cite{zhukov18}.}
	\begin{tabular}{lrr}
		\hline\hline
		Parameter & $n$ type & $p$ type\\
		\hline
		$\wn$ & 70 MHz & 16 MHz\\
		$\wt$ & 16 MHz& 70 MHz\\
		$\tausg$ & 500 ns & 5200 ns\\
		$\taust$ & 10 ns & 35 ns\\
		$\tau_0$ & 0.4 ns & 0.4 ns\\	
		$\TR$  & 13.2 ns & 13.2 ns\\
		$\lambg$ & 1 & 5\\
		$\lambt$ & 5 & 1\\ 
		$g_\mathrm{g}$ & -0.61 & -0.45\\
		$g_\mathrm{t}$ & -0.45 & -0.4\\
		\hline\hline
	\end{tabular}
\end{table}

The calculations are performed for the two sets of parameters summarized in 
Table~\ref{tab:parameters} by applying the two methods described in Appendix~\ref{sec:methods}.
The parameters correspond to $n$-type and $p$-type QD samples~\cite{eh_noise,zhukov18} and illustrate 
the qualitative differences between these two kinds of systems.
The main difference is the strength of the hyperfine interaction $\omega_\mathrm{n}$. 
The electrons are in an $s$-type Bloch band and the hyperfine interaction stems from the Fermi contact interaction, which is strong.
For holes the hyperfine interaction is caused by the dipole-dipole interaction, which is much weaker and anisotropic ($\lambda = 5$)~\cite{testelin07}.
As described above, the situation for the trion state is opposite to that for the ground state.
The longitudinal $g$ factors of electrons and holes are of the same order, but the spin relaxation time in the ground state $\tausg$
is found experimentally to be about one order shorter for electrons than for holes.
The spin relaxation times in the trion states $\taust$ are of the same order for the two types of QDs, but they are much shorter than the spin relaxation times in the ground states.

\subsection{Strong longitudinal magnetic field}
\label{sec:strong_field}

In the limit of a strong longitudinal magnetic field ${\Omega_\mathrm{L} \gg \Omega_\mathrm{N}}$, the hyperfine interaction can be neglected.
Instead of Eqs.~\eqref{eq:dSdt} and~\eqref{eq:dJdt}, the spin dynamics is described approximately by the scalar equations 
\begin{subequations}
	\begin{align}
	\frac{\d S^{z}}{\d t} &= - \frac{S^z}{\tausg} + \frac{J^z}{\tau_0},\\
	\frac{\d J^z}{\d t} &= - \frac{J^z}{\taust} - \frac{J^z}{\tau_0}.
	\end{align}
	\label{eq:Bz}%
\end{subequations}%
In the realistic case $\tausg \gg \TR$, Eq.~\eqref{eq:Bz} can be solved for the initial conditions determined by the pulse relation~\eqref{eq:pulse}.
For now, we only consider resonant pulses ($\Phi = 0$) and find,
for $t\gg\tau_0$ and $t<\TR$,
\begin{equation}
S^z(t)=(S^z_\mathrm{b}+\Delta S^z)\e^{-t/\tausg},
\end{equation}
where
\begin{equation}
\label{eq:Delta_S}
\Delta S^z=-\left(\frac{\mathcal P}{4}+\frac{S^z_\mathrm{b}}{2}\right)(1-Q^2)\left(\frac{\tau_0}{\taust+\tau_0}-\frac{\tau_0}{\tausg}\right)
\end{equation}
is the difference of the spin polarization after the trion recombination and 
before the arrival of the next pump pulse. 
Here, in contrast to Ref.~\cite{smirnov18}, the time $\taust$ can be comparable to $\tau_0$. 

For the same approximation, the train of pump pulses can be replaced on average by a continuous pumping
so that the spin dynamics is described by
\begin{equation}
\label{eq:continuous}
\frac{\d S^z}{\d t}=\frac{\Delta S^z}{\TR}-\frac{S^z}{\tausg},
\end{equation}
where the first term represents the generation rate of the spin polarization.
In this limit, the spin inertia signal is given by~\cite{heisterkamp15}
\begin{equation}
\label{eq:Korenev}
L(\wm)=\frac{L(0)}{\sqrt{1+(\wm\tauseff)^2}},
\end{equation}
where $\tauseff$ is the effective spin relaxation time defined by~\cite{smirnov18}
\begin{equation}
\label{eq:tau_s_eff}
\frac{1}{\tauseff}=\frac{1}{\tausg}+\frac{(1-Q^2)\tau_0}{2\TR(\taust+\tau_0)}.
\end{equation}
For $Q < 1$, this time is shorter than $\tausg$ due to an effective quenching of the spin relaxation time by strong pump pulses. 
This effect can be seen by substituting Eq.~\eqref{eq:Delta_S} in Eq.~\eqref{eq:continuous} and assuming $\tausg \gg \taust$.
The spin inertia signal at zero modulation frequency is determined by the balance between spin generation and spin relaxation
\begin{equation}
\label{eq:L0}
L(0)=\frac{1-Q^2}{2\pi}\frac{\tauseff}{\TR}\left|\frac{\tau_0}{\taust+\tau_0}-\frac{\tau_0}{\tausg}\right|,
\end{equation}
where we took into account that according to the definition~\eqref{eq:L_def} the steady-state spin polarization is $\pi/2$ times larger than $L(0)$~\cite{smirnov18}.
The pump pulse creates spin polarization proportional to $(1-Q^2)$ in the ground and trion states in opposite directions, see Eq.~\eqref{eq:pulse_trion}. 
Hence, the spin relaxation in the ground and trion states during the trion lifetime leads to two contributions to the total spin polarization pointing in opposite directions, 
which is described by the two terms under the modulus.

Note that it follows from Eq.~\eqref{eq:Delta_S} that the effect of spin polarization saturation can be understood also as an effective decrease of the spin generation rate. 
However, it is not straightforward in this interpretation to explain the change of the cut-off frequency $1/\tauseff$ in Eq.~\eqref{eq:Korenev}.
Thus, we prefer the interpretation based on the effective decrease of the spin relaxation time with an increase of the pump power.

\begin{figure*}[bt]
	\centering
	\subfloat[\label{fig:L_fm}]{\includegraphics[width=\columnwidth]{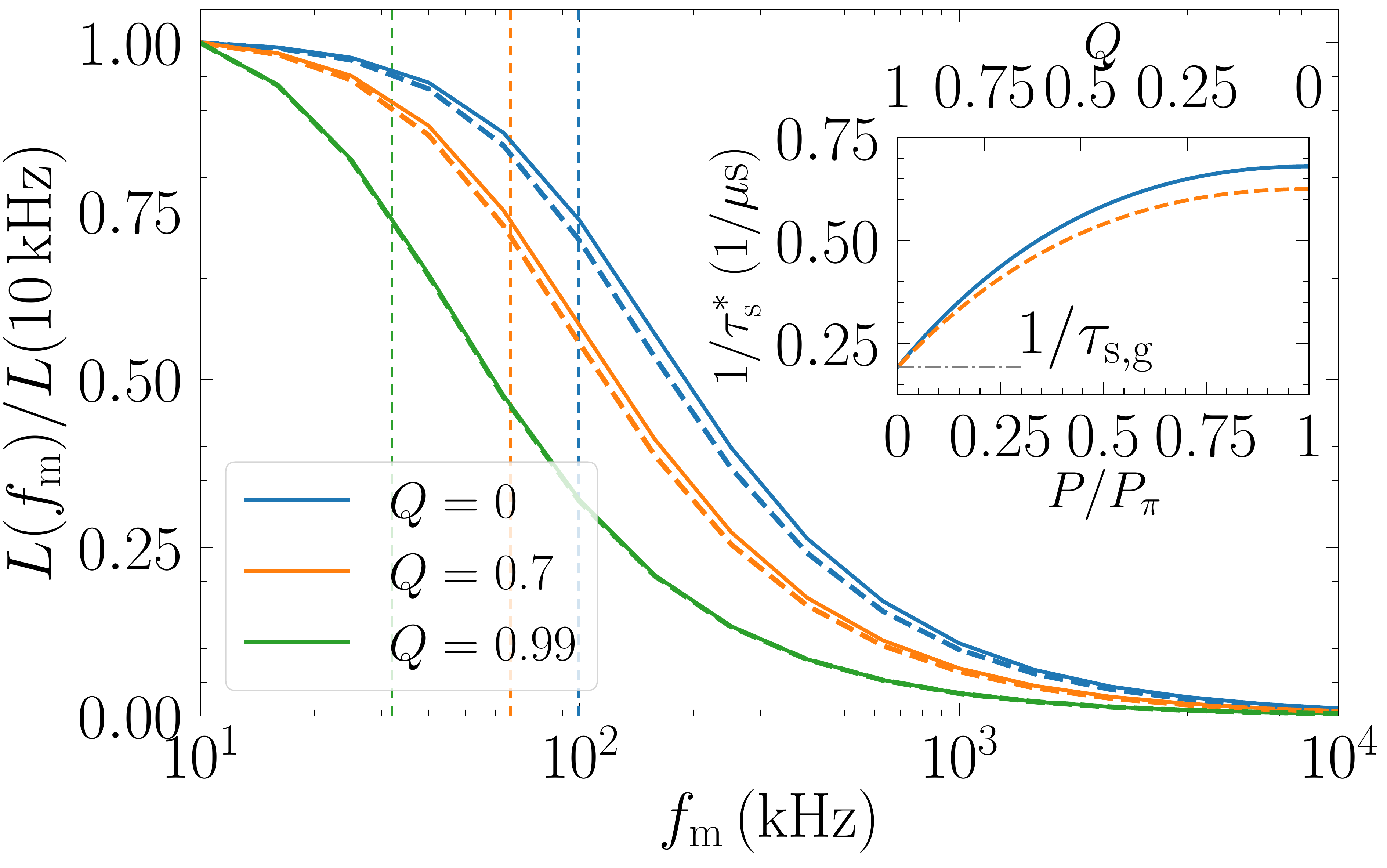}}
	\subfloat[\label{fig:L_P}]{\includegraphics[width=\columnwidth]{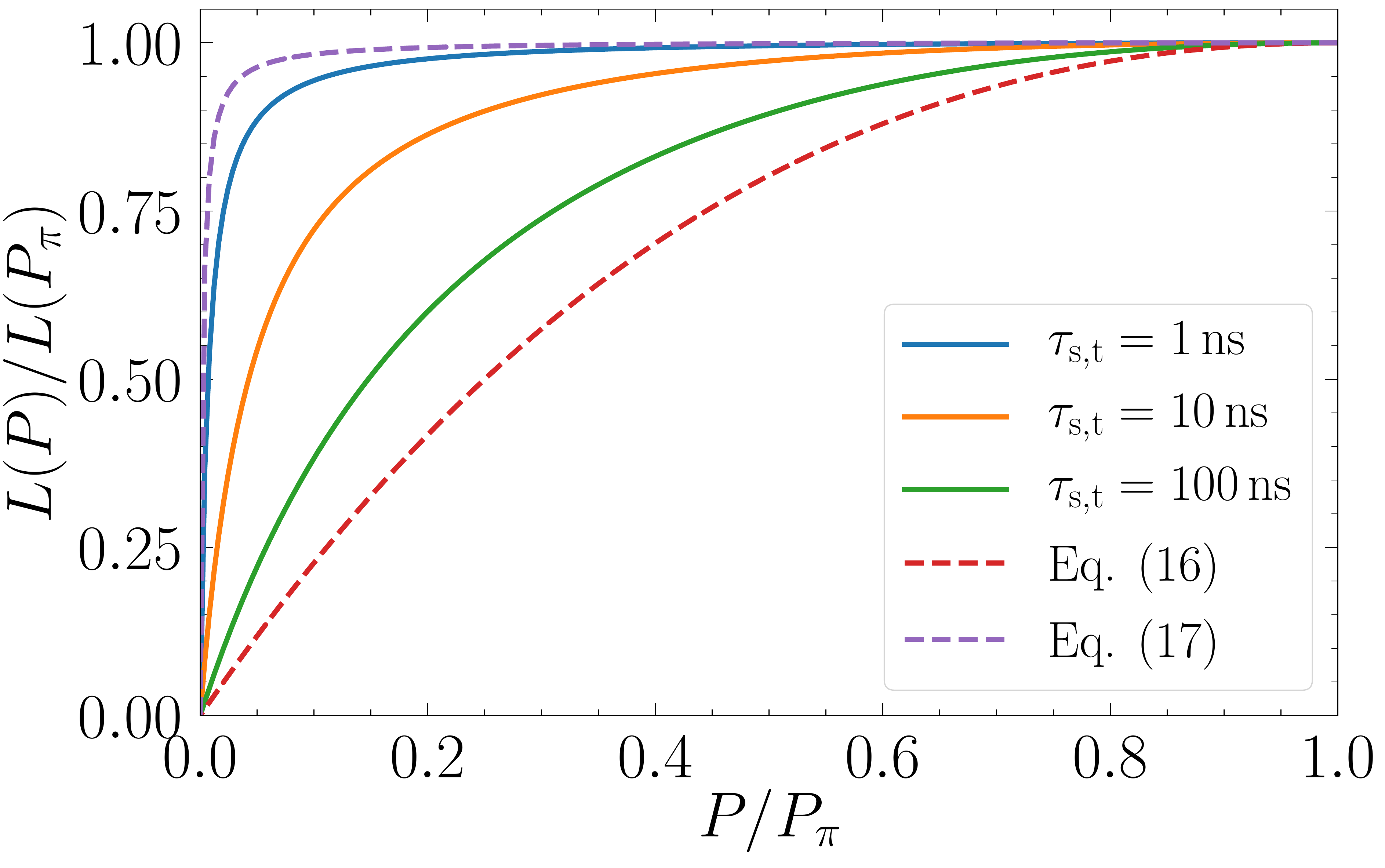}}
	\caption{Analysis of the spin inertia signal $L$ in the limit of a strong longitudinal magnetic field for parameters corresponding to $p$-type samples listed in Table~\ref{tab:parameters}.
		(a) Dependence of the spin inertia signal $L$ on the modulation frequency $\fm$ for fixed magnetic field $\Bext=300$~mT, 
		calculated by numerical simulations (solid lines) and by Eqs.~\eqref{eq:Korenev} and~\eqref{eq:tau_s_eff} (dashed lines).
		The vertical dashed lines represent the corresponding cut-off frequencies $1/2\pi \tauseff$, where $\tauseff$ is calculated according to Eq.~\eqref{eq:tau_s_eff}.
		The inset shows the inverse effective spin relaxation time $1/\tauseff$ as a function of $Q$ and $P / P_\pi$
		calculated by fitting Eq.~\eqref{eq:Korenev} to data obtained from the numerical 
		simulation for $\Bext = 300$~mT (solid line) and calculated according to Eq.~\eqref{eq:tau_s_eff} (dashed line).
		(b) Dependence of the spin inertia signal $L$ on the normalized pump power $P/P_\pi$
		for fixed magnetic field $\Bext = 300$~mT in the steady-state limit 
		$\fm \to 0$ for various trion spin relaxation times $\taust$. 
		For comparison, the limiting cases of large and small $\taust$, described by 
		Eqs.~\eqref{eq:LP1} and~\eqref{eq:LP2}, are depicted as dashed lines.}
	\label{fig:L_fm_P}
\end{figure*}

The spin inertia signal $L(\fm)$ ($\fm=\wm/2\pi$) for a relatively strong longitudinal magnetic field $\Bext=300$~mT is shown in Fig.~\ref{fig:L_fm} for different 
pump powers for $p$-type QDs. 
The spin polarization decreases notably when the modulation frequency $\wm$ becomes larger than $1/\tau_\mathrm{s}^*$, which is the essence of the spin inertia effect.
It is described by Eqs.~\eqref{eq:Korenev} and~\eqref{eq:tau_s_eff}, depicted by the dashed curve in Fig.~\ref{fig:L_fm}. 
The difference between the analytical (dashed lines) and numerical (solid lines) results is related to the large but finite ratio of the external magnetic field and the Overhauser field. 

The result is similar for $n$-type QDs (not shown) since the influence of the hyperfine interaction can be neglected in the limit of a strong longitudinal magnetic field. 
Mainly a shift of the spin inertia signal curve to larger modulation frequencies is found (not shown) since the corresponding ground state spin relaxation time $\tausg$ is shorter by one order of magnitude.

Let us turn to the dependence on the pump power.
For resonant pump pulses ${Q=\cos(\Theta/2)}$ is valid, where $\Theta$ 
is the area of the pump pulse~\cite{yugov09}. 
In parallel, the pump pulse power scales like $P\propto\Theta^2$ \cite{greil06a,yugov12}. Thus, for a power smaller than the power of a $\pi$ pulse,
denoted by $P_\pi$, we obtain the relation
\begin{equation}
Q=\cos\left(\frac{\pi}{2}\sqrt{\frac{P}{P_\pi}}\right),
\label{eq:P_Q}
\end{equation}
which facilitates the investigation of the spin inertia as a function of the pump power $P$ 
because the proportionality factor depends on many details, e.g., on the specific QD sample.

Spin inertia measurements allow for determining the effective spin relaxation time $\tauseff$.
The inset in Fig.~\ref{fig:L_fm} shows the dependence of the inverse effective spin relaxation time on $Q$ and on the pump power $P$, which is related to $Q$ by Eq.~\eqref{eq:P_Q}. 
The solid line is obtained from the numerical simulation by fitting the dependence $L(\wm)$ for $\Bext = 300$~mT with Eq.~\eqref{eq:Korenev}, and the dashed lines are calculated using Eq.~\eqref{eq:tau_s_eff}.
In the limit of zero pump power ($Q \to 1$), $\tauseff$ is equal to the true spin relaxation time $\tausg$, as expected.
With an increase of the pump power, the spin relaxation rate effectively increases.
Its dependence on the pump power in the regime of small pump power appears to be linear. 
This justifies a linear extrapolation to zero pump power in order to extract the intrinsic spin relaxation time $\tausg$ from spin inertia measurements~\cite{zhukov18}.

The dependence of the spin inertia signal on the pump power in the high field limit is shown in Fig.~\ref{fig:L_P} for zero modulation frequency for $p$-type QDs.
Notably, the shape of this dependence is different for different ratios 
$\taust/\tau_0$ of the trion spin relaxation time and the radiative lifetime.
This renders the determination of this ratio possible. 
If the radiative lifetime $\tau_0$ is known, e.g., from time-dependent photoluminescence measurements, the dependence on the pump power gives access to the trion spin relaxation time $\taust$.
Note that in Fig.~\ref{fig:L_P} we vary only $\taust$ while keeping $\tau_0$ constant,
but we checked that the variation of $1/\tau_0$ has the same effect.

Let us analyze $L(0)$ given by Eq.~\eqref{eq:L0}, which describes the spin polarization of the steady state induced by an infinite train of pulses with the same helicity.
Usually, the spin relaxation in the trion state is faster than in the ground state, so
\begin{equation}
\label{eq:L_P}
L(0)=\frac{(1-Q^2)\tauseff\tau_0}{2\pi\TR(\taust+\tau_0)}
\end{equation}
holds. The dependence of spin inertia signal $L(0)$ on the pump power in a large external magnetic field 
stems directly from the multiplier $1-Q^2=\sin^2(\sqrt{P/P_\pi}\pi/2)$ in 
Eq.~\eqref{eq:L_P} and also from the dependence of the effective spin relaxation time 
$\tauseff$ on the pump power. 
For long trion spin relaxation times $\taust\gg\tausg\tau_0/\TR$, the latter 
dependence is negligible and the spin inertia signal as a function of the pump power is given by
\begin{equation}
\label{eq:LP1}
\frac{L(P)}{L(P_\pi)}=\sin^2\left(\frac{\pi}{2}\sqrt{\frac{P}{P_\pi}}\right).
\end{equation}
This dependence is displayed by the red dashed curve in Fig.~\ref{fig:L_P}. 
In the opposite limit $\taust\ll\tau_0$, one finds the dependence
\begin{equation}
\label{eq:LP2}
\frac{L(P)}{L(P_\pi)}=\frac{\sin^2\left(\frac{\pi}{2}
	\sqrt{P/P_\pi}\right)}{\sin^2\left(\frac{\pi}{2}
	\sqrt{P/P_\pi}\right)+2\TR/\tausg},
\end{equation}
where we again use the assumption $\TR\ll\tausg$. 
This dependence is displayed by the purple dashed curve in Fig.~\ref{fig:L_P}. 
In this limit, the spin inertia signal quickly increases for powers $P \gtrsim P_\pi \TR / \tausg$ and then saturates.
This happens because an increase of the pump power not only increases the spin pumping efficiency but also shortens the effective spin relaxation time $\tauseff$
so that both effects compensate each other. 
For intermediate values of $\taust$, the dependence on the pump power of the spin inertia signal smoothly changes from one limit to another as shown in Fig.~\ref{fig:L_P}.
Equation~\eqref{eq:LP2} also indicates that the ground state relaxation time $\tausg$ influences the saturation behavior.
Indeed, when studying the $n$-type parameters for which $\tausg$ is smaller by one order of magnitude, achieving saturation of $L(P)/L(P_\pi)$ requires a larger pump power.

\subsection{Polarization recovery}
\label{sec:polarization_recovery}

Next we analyze the dependence of the spin inertia signal on the strength of 
the external longitudinal magnetic field. It is commonly accepted that the application of 
an external magnetic field in Faraday geometry suppresses nuclei-induced spin relaxation.
Hence, it leads to an increase of the spin polarization~\cite{book_Glazov}. 
This effect is called \emph{polarization recovery}~\cite{heisterkamp15} and the
related dependence is called the polarization recovery curve (PRC). 
Recently, it was shown that this polarization recovery can manifest itself 
in a surprising non-monotonic way in spin inertia measurements~\cite{zhukov18}.

\begin{figure*}[t!]
	\centering
		\subfloat[$n$ type\label{fig:L_Bext_fm_a}]{
		\includegraphics[width=\columnwidth]{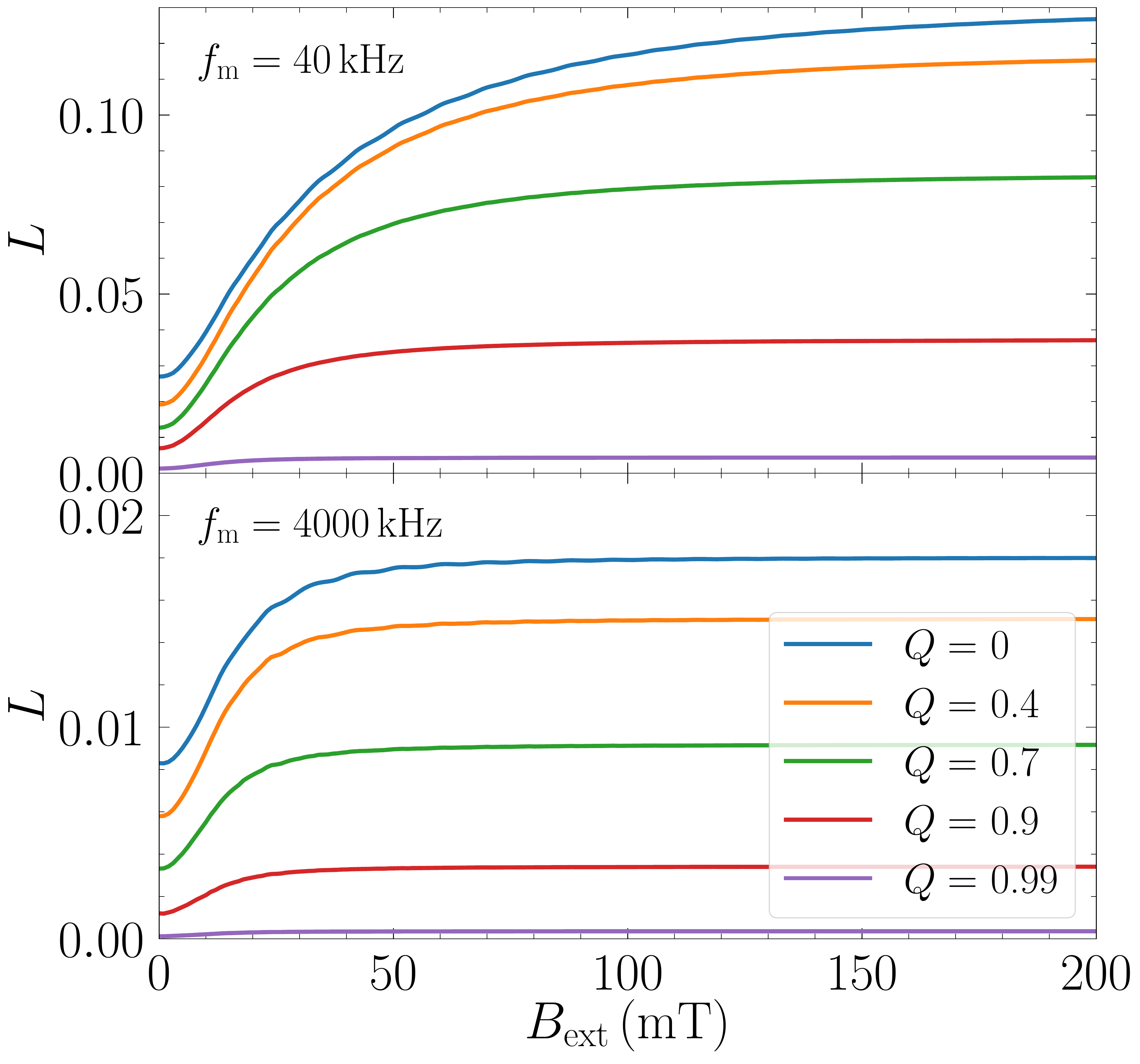}}
		\subfloat[$p$ type\label{fig:L_Bext_fm_b}]{
		\includegraphics[width=\columnwidth]{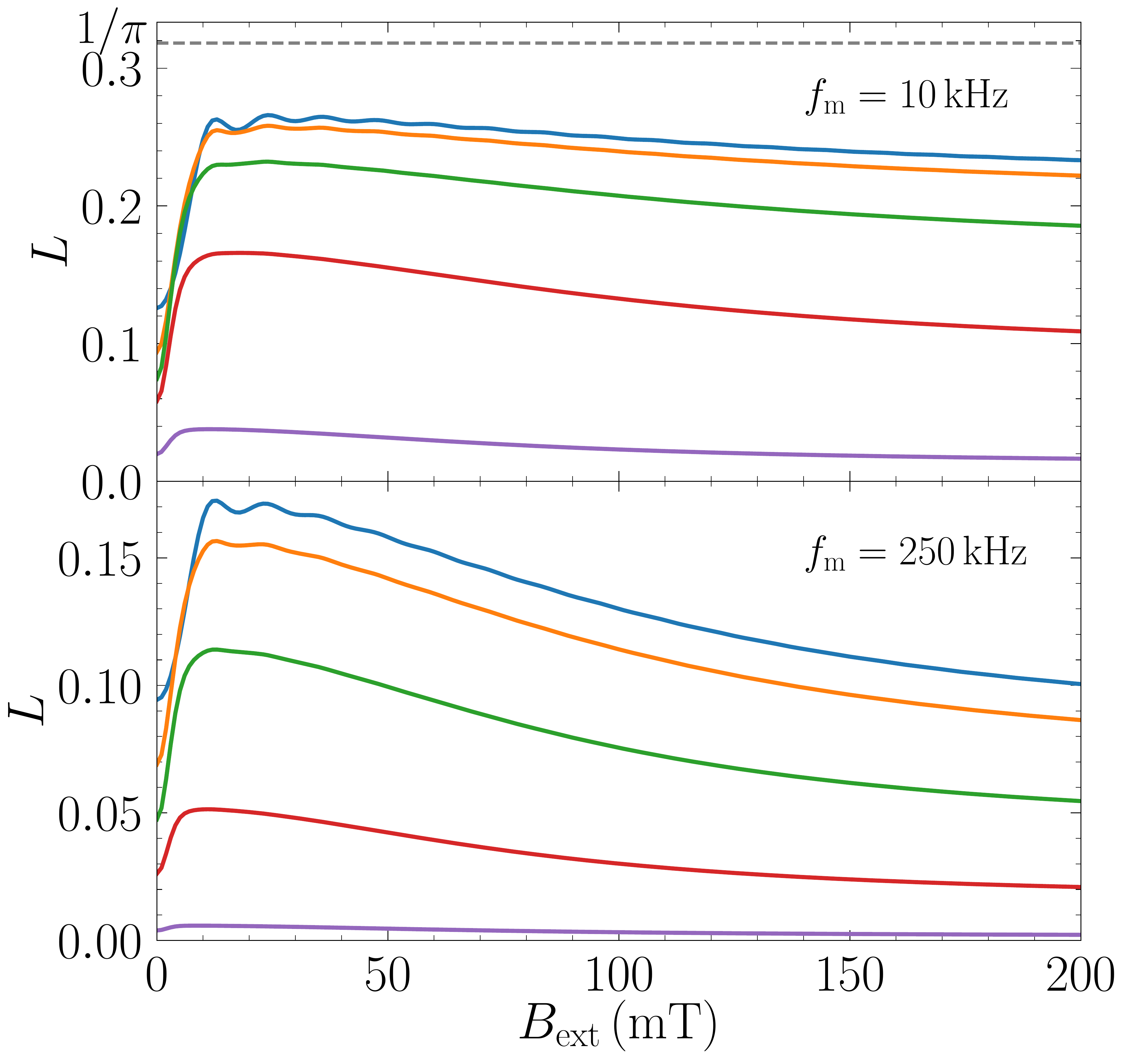}}
	\caption{Polarization recovery curves: spin inertia signal $L$ vs.\
		external magnetic field $\Bext$ for low and high modulation frequencies and for 
		various values of $Q$ as indicated in the plots. 
		The choices of parameters for $n$- and $p$-type QDs are listed in Table~\ref{tab:parameters}. 
		The gray dashed line in panel (b) indicates the maximum possible value $1/\pi$.}
	\label{fig:L_Bext_fm}
\end{figure*}

Figure~\ref{fig:L_Bext_fm} depicts the spin inertia signal as a function of the 
external magnetic field. The calculations are done for $n$-type [Fig.~\ref{fig:L_Bext_fm_a}] and 
$p$-type [Fig.~\ref{fig:L_Bext_fm_b}] QDs for low and high modulations frequencies and for various pumping strengths. 
Obviously, the shape of the PRCs in Fig.~\ref{fig:L_Bext_fm} is qualitatively 
different for the two sets of system parameters: 
For $n$-type QDs it is monotonic; for $p$-type QDs it has a maximum at around $\Bext = 15$~mT. 
Considering that the spin inertia signal is an even function of $\Bext$, these shapes are 
called V-like and M-like, respectively~\cite{smirnov18,zhukov18}. 
We stress that the shape does not change qualitatively with the modulation frequency,
but we find a strong broadening of the PRCs for $n$-type when decreasing the modulation 
frequency as shown in Fig.~\ref{fig:L_Bext_fm_a}.
This suggests that accounting for finite pump power can improve the results of Ref.~\cite{zhukov18} 
where a broadening of the PRC with decreasing modulation frequency was experimentally observed, but not described theoretically.

The origin of the different shapes of the PRCs is the different mechanisms of
spin orientation~\cite{smirnov18}. The polarization generation is related to the 
complete trion spin relaxation mechanism.
For $p$-type QDs, in contrast to $n$-type QDs, a major part of the relaxation stems
from the hyperfine interaction of the unpaired electron spin in the trion with the surrounding nuclei.
Therefore, the application of a longitudinal magnetic field suppresses a part of the 
total trion spin relaxation, which reduces the efficiency of spin polarization generation in the ground state.
As a result, this polarization is reduced for a large magnetic field, leading to the characteristic M-like shape of the PRC for $p$-type QDs.
In contrast, for $n$-type QDs, the hyperfine interaction of the hole in the trion with the surrounding nuclei is weak, so a reduction of the spin polarization in the ground state is not observed, 
leading to the characteristic V-like shape of the polarization recovery.

However, an increase of the pump power can lead to a change of the PRC from M-like to V-like 
as shown in Fig.~\ref{fig:L_Bext_fm_b}.
This is caused by the fact that strong pump pulses can saturate the spin polarization in the ground state even for a reduced spin generation efficiency of a single pulse. 
Hence, the decrease of the total trion spin relaxation rate by an increase of the magnetic field 
does not necessarily lead to a decrease of the spin polarization in the limit of 
strong pumping. The maximally possible value of the spin inertia signal corresponds to the 
case where the spin polarization $S_z=\pm1/2$ is reached before each pump pulse. 
In this limit, Eq.~\eqref{eq:L_def} yields $L=1/\pi$ which is included 
in Fig.~\ref{fig:L_Bext_fm_b} by the gray dashed line as the upper bound.

Another effect related to large pump power is a broadening of the PRC~\cite{yugov09}.
This effect is most clearly seen for $n$-type QDs at low modulation frequency, 
but it is also present in $p$-type QDs. 
A rescaled version of Fig.~\ref{fig:L_Bext_fm_a} is shown in Appendix~\ref{sec:further_results_normalized_PRC} which illustrates this effect even better.
Similarly to the change of the shape of the PRCs, this effect is also related to the 
saturation of the spin polarization. 
Indeed, strong pump pulses partially destroy the spin polarization, as can be seen from Eq.~\eqref{eq:pulse_Sz} for $Q \to 0$. 
This leads to an effective decrease of the spin polarization in a large magnetic field, 
which in turn induces a broadening of the PRCs.

Notably, Fig.~\ref{fig:L_Bext_fm} shows that the pump power affects the 
amplitude of the dip at zero magnetic field.
This effect is similar to what is observed when including slow nuclear spin dynamics in the model~\cite{smirnov18} so
that the finite nuclear spin correlation time obtained in Ref.~\cite{zhukov18} may be related to the assumption of weak pumping.

The influence of a large pump power is partially suppressed for $n$-type QDs for high 
modulation frequency, see Fig.~\ref{fig:L_Bext_fm_a}.
This effect becomes much more conspicuous if the curves are scaled on one another, see Appendix~\ref{sec:further_results_normalized_PRC} for more details.
Therefore, we suggest that the measurement of the PRCs at high modulation frequency enables
a better determination of the spin dynamics parameters, especially those related to the 
hyperfine interaction.
Essentially, one has to stay away from the saturation limit of the spin polarization, which is achieved by applying weak pulses and measuring at high modulation frequencies.

\subsection{Non-resonant pumping}

\begin{figure}[tb!]
	\centering
		\subfloat[$n$ type\label{fig:L_Phi_Q_a}]{
		\includegraphics[width=\columnwidth]{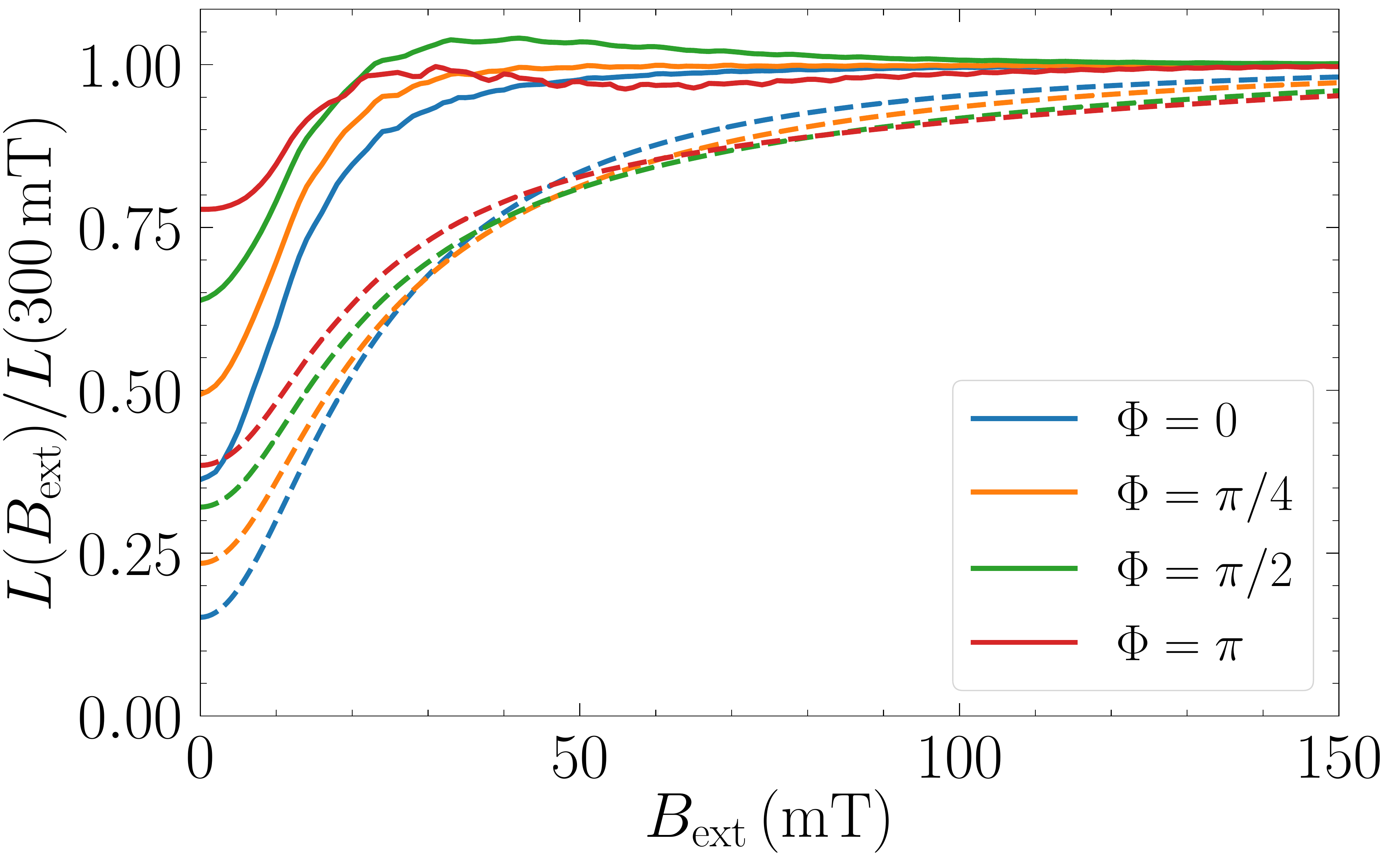}}\\
		\subfloat[$p$ type\label{fig:L_Phi_Q_b}]{
		\includegraphics[width=\columnwidth]{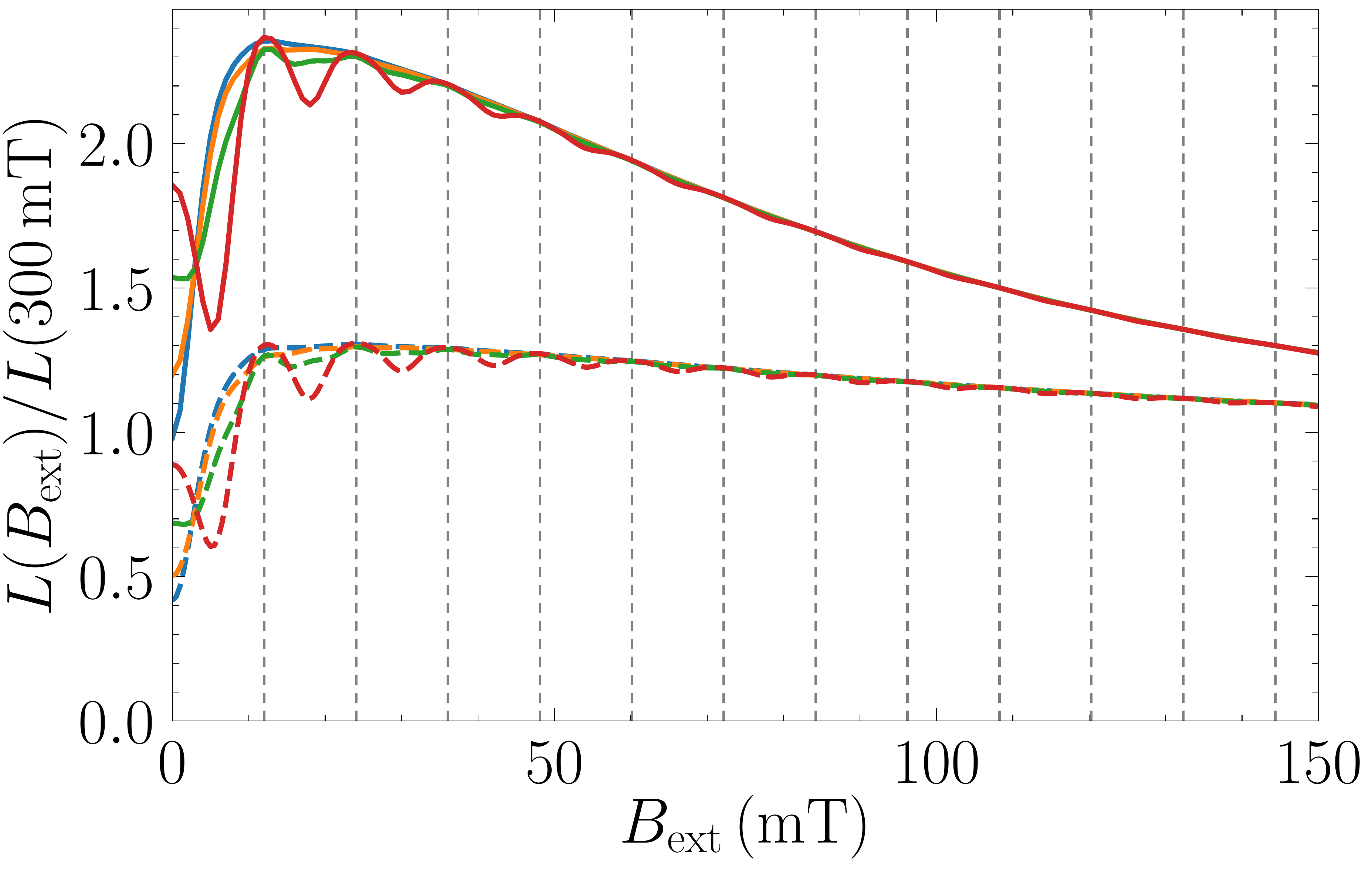}}
	\caption{Polarization recovery curves for moderate pump pulses ($Q=0.7$) and various values of $\Phi$. 
		The numerical simulations (solid lines) are performed for $\fm = 4000$~kHz ($n$ type) and $\fm = 250$~kHz ($p$ type) and for $\fm=0$ (dashed lines).
		The vertical dashed lines in panel (b) represent the phase synchronization condition~\eqref{eq:ML_ensemble}.
		The choices of parameters for $n$- and $p$-type QDs are listed in Table~\ref{tab:parameters}.}
	\label{fig:L_Phi_Q}
\end{figure}

Up to now, we have only considered resonant pumping of the QDs by assuming 
$\Phi=0$ in the pulse relation~\eqref{eq:pulse}. Let us briefly discuss the role of detuned pulses
with $\Phi \ne 0$. 
Figure~\ref{fig:L_Phi_Q} shows the PRCs for $n$-type [Fig.~\ref{fig:L_Phi_Q_a}] and $p$-type [Fig.~\ref{fig:L_Phi_Q_b}] QDs 
for the two different values of $Q$ and for a series of values of the phase $\Phi$.
There are a number of changes in the PRCs for increasing $\Phi$, but one can see that the 
PRCs are qualitatively similar for large (solid lines) and 
zero (dashed lines) modulation frequency.

In the first place, the increase of $\Phi$ leads to an increase of the spin polarization 
for small to intermediate magnetic fields.
When studying the unscaled versions of the PRCs in Fig.~\ref{fig:L_Phi_Q} (not shown), we find that the spin polarization does not depend on $\Phi$ at large magnetic fields.
To explain this effect we recall that a finite value of $\Phi$ corresponds to a spin rotation 
in the $(xy)$ plane as described by the pulse relation~\eqref{eq:pulse}. 
This is equivalent to an additional longitudinal magnetic field acting only during the 
pump pulse. 
The spin rotation is caused by the effective additional longitudinal magnetic field
appearing due to the dynamic Zeeman effect, which suppresses the role of nuclear fluctuations~\cite{Kimel2005,OpticalField}.
The dynamic Zeeman effect results from the optical Stark effect in the field of the circularly polarized light.
Qualitatively, this additional magnetic field increases the spin polarization 
similarly to the real external magnetic field.

Moreover, strong pump pulses partially erase the in-plane spin polarization, 
see Eq.~\eqref{eq:pulse} for $Q \to 0$. 
As a result, the in-plane spin rotations are less important for the stronger pulses and 
thus, the increase of the spin inertia signal for increasing $\Phi$ is smaller 
for smaller $Q$ (not shown). 

In order to describe the effect of detuning, i.e., finite $\Phi$, on the spin polarization
analytically, let us consider the limit of short trion spin relaxation time 
$\taust \ll \tau_0$ for isotropic hyperfine interactions
in the ground state $\lambg = 1$, long spin relaxation time $\tausg \gg \TR$, strong hyperfine interaction 
$\omega_\mathrm{n,g} \TR \gg 1$, and weak pulses $(1-Q^2)\tausg/\TR \ll 1$. The hyperfine interaction in the trion state plays no role in this limit. 
Under these assumptions, the system of equations~\eqref{eq:pulse} 
and~\eqref{eq:Smp} can be solved analytically for the steady-state solution yielding
\begin{equation}
\label{eq:Szb_detuning}
S^z_\mathrm{b} = -\frac{\mathcal{P}}{4}(1-Q^2)\frac{\tausg}{\TR} \frac{
	 \kappa^2
}{
	\mathcal \kappa^2 + \sin^2(\theta) \sin^2(\phi/2)
},
\end{equation}
where
\begin{equation}
	\kappa = \cos(\theta)\cos(\Phi/2)\sin(\phi/2)- \mathcal{P} \cos(\phi/2)\sin(\Phi/2)
\end{equation}
with $\phi = \Omega_\geff \TR$, 
$\bm\Omega_\geff=\bm\Omega_\mathrm{N,g}+\bm\Omega_\mathrm{L,g}$, and $\theta$ 
is the angle between $\bm{\Omega}_\geff$ and the $z$ axis. 
Then, averaging this result over $\bm\Omega_\mathrm{N,g}$ with its distribution~\eqref{eq:distribution} yields the ratio
\begin{multline}
\frac{L(\Bext = 0)}{L(\Bext = \infty)} = 
1 - \frac{2}{\sin^2(\Phi/2)} \\ + \frac{2\pi \sin^4(\Phi/4) + 
	|\Phi| \cos(\Phi/2)}{\sin^3(|\Phi|/2)},
\label{eq:L0_Linfty}
\end{multline}
where we assume $|\Phi| < \pi$.
In the particular cases of $\Phi = 0$ and $\pi$, we find ${L(\Bext=0)/L(\Bext=\infty) = 1/3}$ 
and $\pi/2 - 1$, respectively, with a monotonic increase inbetween.
In conclusion, we find that a finite phase $0 < \Phi \le \pi$ leads to an increase of the 
spin inertia signal at zero magnetic field relative to its value at large magnetic field.

A more general numerical analysis is presented in Appendix~\ref{sec:further_results_detuning}.
It reveals a flattening of the $\Phi$ dependence described by Eq.~\eqref{eq:L0_Linfty} when $Q$ is decreased.
In agreement with Eq.~\eqref{eq:L0_Linfty}, an increase of $\Phi$ leads to a generalization of the established 
classical ratio~$1/3$~\cite{merku02,petrov08} of the spin polarization at zero field relative to the polarization at large magnetic field.

The most interesting effect of detuning is a qualitative change of the shape of the 
PRCs upon increasing $\Phi$. 
For $n$-type QDs, the PRC becomes non-monotonic while for $p$-type QDs 
one finds additional oscillations caused by RSA, see below.

\subsection{Resonant spin amplification in Faraday geometry}
\label{sec:RSA}

In Fig.~\ref{fig:L_Phi_Q_b} oscillations in the spin polarization as a function of
the longitudinal external magnetic field are visible. 
These oscillations are a manifestation of RSA in Faraday geometry.
We recall that RSA is well established in Voigt geometry
where a transverse external magnetic field is applied~\cite{Kikkawa98,yugov12,varwig12,zhukov12}. 
It consists in a considerable enhancement of the spin polarization if the spin 
precession frequency matches the phase synchronization condition (PSC)
\begin{equation}
\Omega_\mathrm{L,g}\TR=2\pi k,
\label{eq:ML_ensemble}
\end{equation}
where $k$ is an integer number.
If this condition is fulfilled, the spin of the charge carrier precesses an integer number of 
times around the magnetic field between the pump pulses
leading to constructive interference between the spin polarization created by consecutive pulses.

\begin{figure}[b!]
	\centering
	\includegraphics[width=0.8\columnwidth]{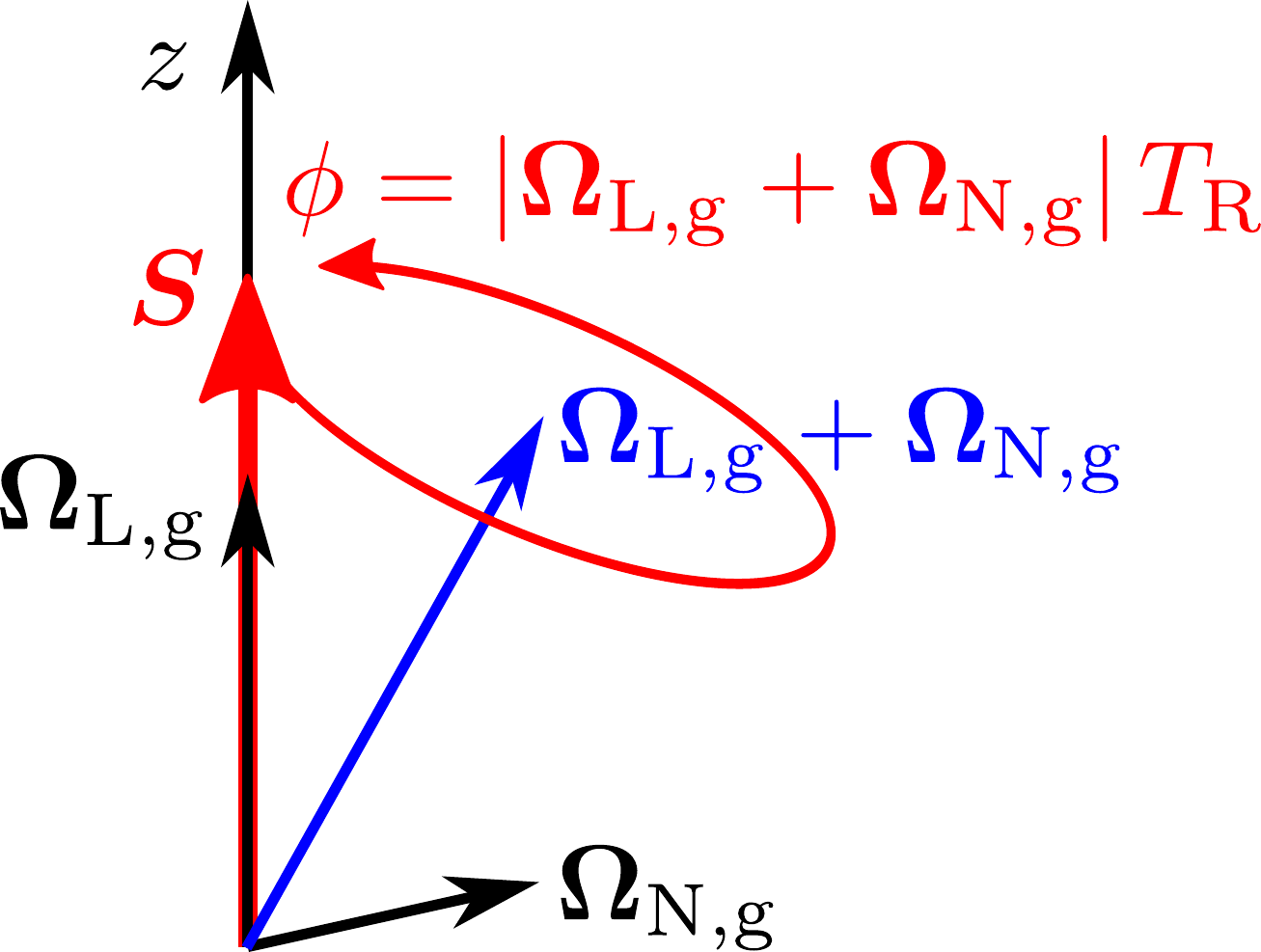}
	\caption{Illustration of the mechanism leading to the emergence of RSA due to the transverse components of the Overhauser field $\bm \Omega_\mathrm{N,g}$, which tilt the total spin precession frequency from the $z$ axis.}
	\label{fig:ML_sketch}%
\end{figure}

In Ref.~\cite{zhukov12} it was demonstrated that applying a magnetic field which is slightly tilted from Faraday geometry reveals RSA peaks in PRC measurements.
However, in the case under study, the external magnetic field is applied along the $z$ axis without any tilt, so the spin component $S^z$ does not precess on average and one may argue that RSA cannot be observed. 
However, the Overhauser field slightly tilts the total spin precession frequency from the 
$z$ axis, see Fig.~\ref{fig:ML_sketch}, which renders RSA observable if 
$\Omega_\mathrm{N,g}$ and $\Omega_\mathrm{L,g}$ are of similar value. 
This effect emerges due to the oscillations in the time domain of an initially created spin polarization in QDs subjected to a longitudinal magnetic field~\cite{merku02}.
The random Overhauser field smears out the RSA peaks
so that they can disappear, for instance, for the $n$-type parameters where 
$\wn$ is significantly larger.
Only in Fig.~\ref{fig:L_Bext_fm_a} can one discern very tiny oscillations 
for $Q = 0$ which can be attributed to RSA.

Resonant spin amplification in Faraday geometry can be observed when the condition
$\wn \lesssim \sqrt{2}\pi/\TR$ is fulfilled; otherwise potential RSA peaks are too broad.
This condition is equal to the condition given in Ref.~\cite{yugov12}, where the parameter regimes for RSA and spin mode locking in Voigt geometry are established.
Accordingly, decreasing the repetition time $\TR$ of the pump pulses in 
measurements of $n$-type samples will reveal RSA peaks
and will enhance their appearance in measurements of $p$-type samples.
For the parameters studied in this paper, we have checked numerically that halving the repetition time 
is sufficient to make RSA appear clearly.

\begin{figure}[t!]
	\centering
	\includegraphics[width=\columnwidth]{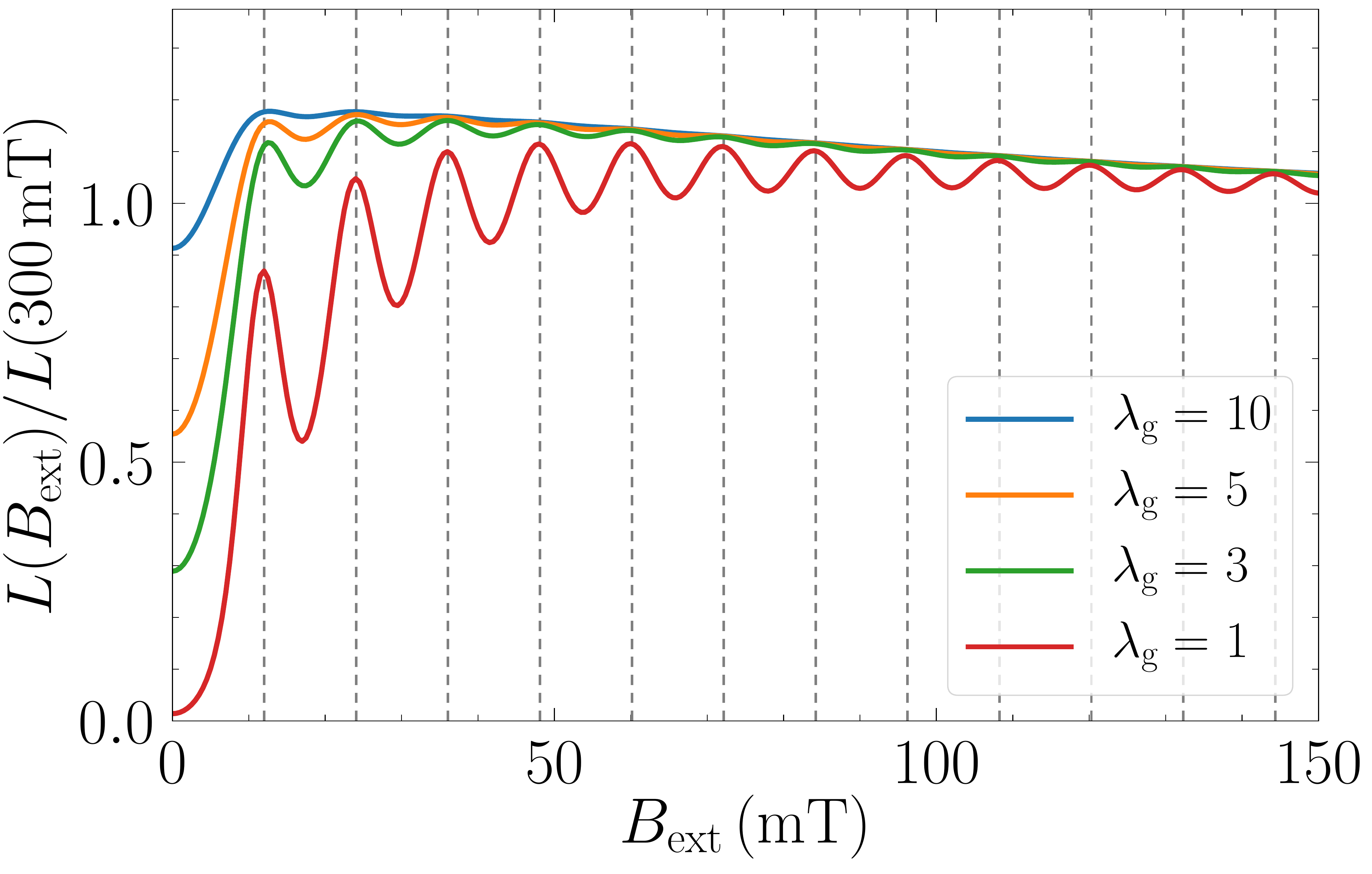}
	\caption{Polarization recovery curves for $p$-type samples for various anisotropy degrees $\lambg$ in the limit of zero modulation frequency $\fm=0$ for $\pi$ pulses ($Q=0$).
		The other parameters are chosen according to Table~\ref{tab:parameters}. 
		The dashed vertical lines represent the values of $\Bext$ which fulfill the PSC~\eqref{eq:ML_ensemble}.}
	\label{fig:PRC_RSA}%
\end{figure}

As explained above, the RSA peaks
result from the transverse components of the Overhauser field $\Omega_\mathrm{N,g}^{x,y}$.
If these components are small due to large anisotropy factors $\lambg$, the amplitude of the RSA peaks is reduced.
Figure~\ref{fig:PRC_RSA} shows the spin inertia signal in the limit of zero modulation frequency for $\pi$ pulses ($Q = 0$)
for various anisotropies of the hyperfine interaction $\lambg$, see Eq.~\eqref{eq:distribution}. 
For comparison, vertical dashed lines represent the values of $\Bext$ which fulfill the PSC~\eqref{eq:ML_ensemble}.
The maxima of the oscillations match the PRC perfectly. 
Only a tiny shift can be discerned for the first maximum, which is expected for small values of $\Bext$ according to Eq.~\eqref{eq:ML_deviation}.
As expected, smaller values of $\lambg$ are favorable to observe RSA peaks because the transverse components of the Overhauser field are less suppressed.

A certain deviation from the PSC~\eqref{eq:ML_ensemble} 
is expected for small magnetic fields. 
For each individual quantum dot of the ensemble the relevant resonance condition reads
\begin{equation}
\Omega_\geff \TR = |\bm\Omega_\mathrm{L,g} + \bm\Omega_\mathrm{N,g} |\TR=2\pi k
\label{eq:ML_individual}
\end{equation}
because the Larmor precession takes place around the complete magnetic field
consisting of the external one and the Overhauser field. Assuming that the
external field is nevertheless larger than the characteristic value $\wn$ of the Overhauser 
field one can expand the modulus in Eq.~\eqref{eq:ML_individual} in powers of the Overhauser field
and average the resulting expression over the distribution~\eqref{eq:distribution},
yielding
\begin{equation}
\overline{|\bm\Omega_\mathrm{L,g} + \bm\Omega_\mathrm{N,g} |}  = 
\Omega_\mathrm{L,g}  \left[1 + \frac{\wn^2}{2 \lambg^2 \Omega_\mathrm{L,g}^2} +
\mathcal{O}\left(\frac{\wn^4}{\Omega_\mathrm{L,g}^4}\right)\right].
\label{eq:ML_deviation}%
\end{equation}
Contributions from odd orders cancel due to the symmetry of the distribution~\eqref{eq:distribution}.
This qualitatively implies that the resonance condition refers to a slightly higher frequency than given by the bare external magnetic field.
This slight deviation can be discerned around the first resonance in Fig.~\ref{fig:PRC_RSA}.

Resonant spin amplification is less pronounced for weaker pulses corresponding to larger $Q$. 
Strong pump pulses erase the $x$ and $y$ components of the spin as discussed above (see Eq.~\eqref{eq:pulse}),
effectively pushing the spin polarization towards the $z$ axis.
This leads to a decrease of the spin polarization in the QDs where the Overhauser field does not satisfy the PSC~\eqref{eq:ML_individual}.
As a result, the RSA peaks are more pronounced for stronger pump pulses. 
This probably explains why it was not observed in the experiment reported in 
Ref.~\cite{zhukov18} where weak pump pulses were applied.

The assumption of a static Overhauser field does not restrict the observation of RSA in Faraday geometry. In fact, the typical timescale of the Overhauser field dynamics $\tau_\mathrm{c} \approx 200$~ns \cite{zhukov18} is much larger than the pulse repetition time $\TR$, leading to no noticeable smearing of the RSA peaks.

The observation of RSA in Faraday geometry opens the possibility to measure the longitudinal $g$ factor of the charge carriers in small magnetic fields.
This parameter is hardly accessible by other means, so this finding is of high practical importance. 

Another promising outlook based on longitudinal RSA is the possibility to study whether nuclear frequency focusing can emerge in Faraday geometry, but including the dynamics of the Overhauser field is required.
This effect is by now well established in transversal Voigt geometry 
\cite{greil07a,greil09,barnes11,petro12,glazo12a,economou14,beuge16,beuge17,jasch17,scher18,klein18}, but so far neither theoretically predicted nor experimentally
observed in longitudinal Faraday geometry. 
The essential mechanism is commensurability of internal dynamics with the periodicity of external driving by pump pulses. 
The observation of longitudinal RSA in our numerical and analytical
calculations reveals the importance of commensurability also in Faraday geometry.
Hence, the longitudinal RSA effect suggests that
it can be enhanced by nuclear frequency focusing, calling for further investigations.

\section{Conclusions}
\label{sec:conclusions}

We developed a theory of the spin inertia effect for localized charge carriers in singly charged quantum dots,
taking into account a finite pump power and a finite detuning between 
the pump frequency and the trion resonance frequency. 
In strong longitudinal magnetic field, the spin dynamics can be described by a single effective spin relaxation time which shortens for increasing pump power. 
The dependence of the spin inertia signal on the pump power allows one to determine 
the ratio of the trion spin relaxation time and the trion lifetime. 

The dependence of the spin inertia signal on the longitudinal magnetic field, called the
polarization recovery curve, can be V-like or M-like for the parameters of the 
spin dynamics generic for $n$-type and $p$-type quantum dots, respectively. 
An increasing pump power leads to a gradual change of the shape from 
M-like to V-like for low modulation frequencies. 
Large pump power and small modulation frequencies also increase the width of the PRC.
These changes of the PRCs are caused by the saturation of the spin polarization.

Furthermore, we find that detuning the frequency of the pump pulses away from 
the trion resonance can lead to an increase of the spin polarization created 
by the infinite train of pump pulses at zero magnetic field 
relative to its value at large magnetic field. 
Notably, this generalizes the so far established classical ratio~$1/3$ 
of the spin polarization at zero field relative to the polarization at large magnetic field.

Finally and perhaps most importantly, we predict the emergence of resonant spin amplification in Faraday geometry.
This effect manifests itself as oscillations of the spin polarization as a function 
of the external magnetic field in Faraday geometry. It is a clear identifying characteristic of
the commensurability of the Larmor precession with the periodic pumping.
Based on this observation, we suggest to study whether nuclear frequency focusing also occurs in Faraday geometry.

\begin{acknowledgements}
We thank Mikhail M.\ Glazov and Oleg P.\ Sushkov for many helpful discussions.
The authors P.S. and G.S.U. gratefully acknowledge the computing time granted on the 
supercomputer Hazel Hen at the High Performance Computing Center Stuttgart under the project NFFinQDs.
G.S.U. acknowledges the hospitality of the UNSW, Sydney, where this paper was partially written, and the support by the Heinrich-Hertz Stiftung, which helped to make this stay possible.
D.S.S. was partially supported by the Basis Foundation, RF President Grant No. MK-1576.2019.2 
and by the Russian Foundation for Basic Research (Grant No. 17-02-0383). 
This study has been supported financially by the Deutsche Forschungsgemeinschaft and 
the Russian Foundation for Basic Research (Grant No. 19-52-12038) in 
the International Collaborative Research Centre TRR 160 (Projects No.~A4 and No.~A7).
\end{acknowledgements}

\begin{appendix}
\section{Methods}
\label{sec:methods}

For the numerical calculations of the spin inertia signal, we apply the following two approaches: 
the full solution of the equations of the spin dynamics for finite modulation frequency 
and a simplified approach for zero modulation frequency.

\subsection{Finite modulation frequency}
\label{sec:methods_finite_fm}

In order to calculate the spin inertia signal defined by Eq.~\eqref{eq:L_def} for a finite 
modulation frequency $\wm$, we solve the equations of motion~\eqref{eq:dSdt} and~\eqref{eq:dJdt} 
numerically by applying the Dormand-Prince method provided by 
\emph{odeint}~\cite{ahnert11}. The subroutine is part of the Boost C++ library.
The numerical calculation of the spin dynamics is performed for $10^5$ random initial conditions of the Overhauser field 
$\bm \Omega_\mathrm{N,g}$ which are sampled from the Gaussian distribution~\eqref{eq:distribution}.
All components of the vectors $\S$ and $\bm J$ are set equal to zero before the arrival of the
first pump pulse.
Then, periodic pulsing with repetition time $\TR$ is implemented in the simulation 
by solving the equations of motion on a grid given by the intervals 
$t \in [k \TR,\, (k + 1) \TR]$ and applying the pulse~\eqref{eq:pulse} in between.
The helicity of the pulse is switched between $\sigma^+$ and $\sigma^-$ by switching 
the sign of $\mathcal{P}$ in the pulse relation~\eqref{eq:pulse} every $T_\mathrm{m}/2 = \pi/\wm$.
Finally, $S^z(t)$ is averaged over all configurations of the Overhauser field.
The numerical effort can be realized efficiently by massive parallelization.

Since it is unfeasible to perform more than a few simulations of up to around $10^7$ pulses, we
investigate the convergence of $L$ on the number of modulation periods $n_\mathrm{period}$.
We find that simulating only $n_\mathrm{period} = 5$ modulation periods and 
calculating $L$ over the last two periods to exclude effects of transient behavior 
is sufficient to achieve a relative error of $L$ which is well below 1\%. 
For high modulation frequencies this error can grow to about 4\% due to 
the generic non-commensurability of $\Tm$ and $\TR$ which we do not account for.
However, this issue only arises in the case without external magnetic field;
for finite magnetic field the error is smaller.

We note that in the experiment of Ref.~\cite{zhukov18} the delay between the 
pump and the probe pulses was $\tau_\mathrm{d} = -50$~ps.
We checked that this small delay has only a negligible influence on the results 
compared to $\tau_\mathrm{d}=-0$, which we use in our theoretical considerations.

\subsection{Zero modulation frequency}
\label{sec:methods_zero_fm}

In the limit of zero modulation frequency parts of the calculations can be performed 
analytically. In this limit, the spin inertia signal $L$ tends to $2S^z_\mathrm{b}/\pi$ 
as follows from Eq.~\eqref{eq:L_def}. The analytical integration of Eq.~\eqref{eq:dJdt} 
yields the trion spin dynamics. The dynamics of the $z$ component of the trion spin
after a pump pulse is given by
\begin{equation}
\label{eq:Tz}
J^z(t) = J^z_\mathrm{a} [\cos^2(\theta_\mathrm{t}) + \sin^2(\theta_\mathrm{t}) 
\cos(\Omega_\mathrm{eff,t} t)] \e^{-t/\taust} \e^{-t/\tau_0}.
\end{equation}
Here, $\bm{\Omega}_\mathrm{eff,t} = \bm\Omega_\mathrm{N,t}+\bm\Omega_\mathrm{L,t}$ is the 
total trion spin precession frequency and 
$\theta_\mathrm{t}$ is the angle between $\bm{\Omega}_\mathrm{eff,t}$ and 
the $z$ axis.

The spin dynamics in the ground state between two consecutive pulses is determined 
by Eq.~\eqref{eq:dSdt}. In the steady state, its solution at time $t = \TR$ must coincide with $\bm S_\mathrm{b}$~\cite{yugov09,greil06b,shaba03}, leading to the equations
\begin{widetext}
\begin{subequations}
	\begin{multline}  
	S^x_\mathrm{b}=S^x_\mathrm{a}\left[\sin^2(\theta)+\cos^2(\theta)
	\cos(\Omega_\geff \TR)\right]\e^{-\TR/\tausg} 
	-S^y_\mathrm{a}\cos(\theta) \sin(\Omega_\geff \TR)\e^{-\TR/\tausg}  \\
	+\int\left[S^z_\mathrm{a}\delta(\tau)+\frac{J^z(\tau)}{\tau_0}\right]
	\cos(\theta)\sin(\theta)
	\left\{1-\cos\left[\Omega_\geff(\TR-\tau)\right]\right\}
	\e^{-(\TR-\tau)/\tausg}\d \tau,
	\end{multline}
	\begin{multline}
	S^y_\mathrm{b}=S^x_\mathrm{a}\cos(\theta)\sin(\Omega_\geff \TR)
	\e^{-\TR/\tausg} 
	+S^y_\mathrm{a}\cos(\Omega_\geff \TR)\e^{-\TR/\tausg}  \\
	-\int\left[S^z_\mathrm{a}\delta(\tau)+\frac{J^z(\tau)}{\tau_0}\right] 
	\sin(\theta) \sin\left[\Omega_\geff(\TR-\tau)\right]\e^{-(\TR-\tau)/\tausg}\d \tau,
	\end{multline}
	\begin{multline}
	S^z_\mathrm{b}=S^x_\mathrm{a}\cos(\theta)\sin(\theta)
	\left[1-\cos(\Omega_\geff \TR)\right]\e^{-\TR/\tausg} 
	+ S^y_\mathrm{a}\sin(\theta)\sin(\Omega_\geff \TR)\e^{-\TR/\tausg} \\
	+\int\left[S^z_\mathrm{a}\delta(\tau)+\frac{J^z(\tau)}{\tau_0}\right] 
	\left\{\cos^2(\theta)+\sin^2(\theta)\cos\left[\Omega_\geff(\TR-\tau)\right]\right\}
	\e^{-(\TR-\tau)/\tausg}\d \tau.
	\end{multline}
	\label{eq:Smp}%
\end{subequations}
\end{widetext}
Here, $\theta$ is the angle between $\bm{\Omega}_\geff$ and the $z$ axis and without loss of generality we assume that both $\bm{\Omega}_\geff$ and $\bm{\Omega}_\teff$ lie in the 
$(xz)$ plane. 
We also assume that $\tau_0$ is shorter than $\tausg$ and $J^z(\tau)=0$ at $\tau<0$,
so the limits of integration can be extended to~$-\infty$~and~$+\infty$. 
Substitution of Eq.~\eqref{eq:Tz} allows us to solve the integrals analytically, 
but the result is cumbersome. 
Finally, the solution of the coupled set of linear equations~\eqref{eq:pulse} 
and~\eqref{eq:Smp} yields the spin polarization in the steady state.

In order to correctly obtain the limit of zero modulation frequency, 
the solution has to be averaged over the two helicities $\mathcal{P}=\pm1$. 
Technically, this is equivalent to averaging over two opposite detunings 
corresponding to opposite signs of $\Phi$. 
Thus, this averaging is only required for $\Phi \ne 0$.

Finally, averaging of $S^z_\mathrm{b}$ in the steady state over the distribution of the Overhauser field~\eqref{eq:distribution} yields the spin inertia signal
as a function of various parameters in the limit of zero modulation frequency $\fm \to 0$.
Note that a similar approach was used in Refs.~\cite{greil06a,greil06b,greil07a}.

\begin{figure}[b!]
	\centering
	\includegraphics[width=\columnwidth]{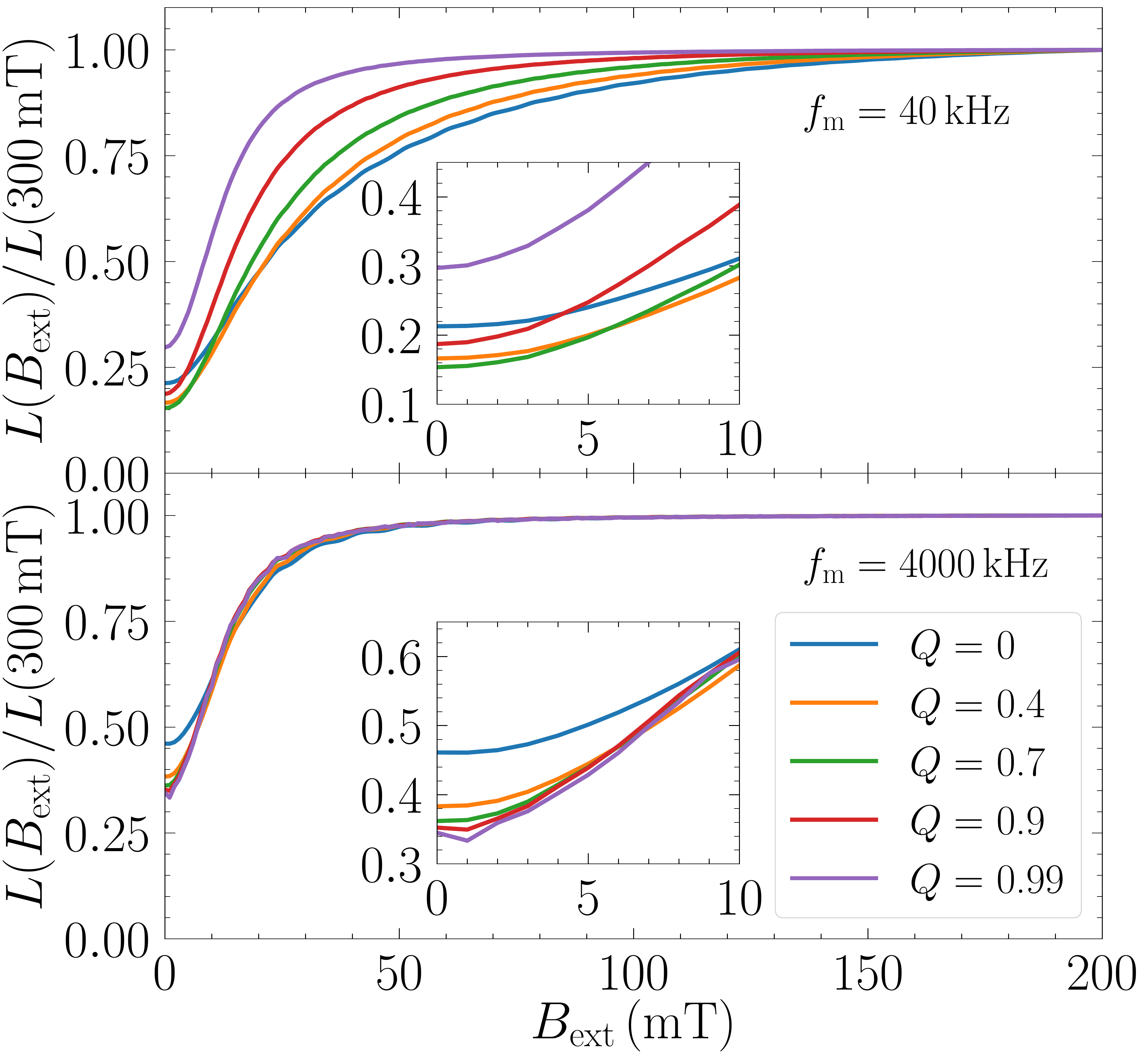}
	\caption{Magnetic field dependence of the \emph{normalized} spin inertia signal $L(\Bext)/L(300\,\mathrm{mT})$ for low and high modulation frequencies and for 
		various values of $Q$ as indicated in the plots. 
		The chosen system parameters correspond to $n$-type QDs listed in Table~\ref{tab:parameters}. This figure is a rescaled version of Fig.~\ref{fig:L_Bext_fm_a}.}
	\label{fig:L_Bext_fm_rescaled}%
\end{figure}

\section{Complementary results}
\label{sec:further_results}

\subsection{Normalized PRC for $n$-type QDs}
\label{sec:further_results_normalized_PRC}

Changes in the shape of PRCs are easier to discern when the curves are normalized, e.g., by rescaling them.
Figure~\ref{fig:L_Bext_fm_rescaled} shows a normalized version of Fig.~\ref{fig:L_Bext_fm_a} by plotting the ratio $L(\Bext)/L(300\,\mathrm{mT})$.
The upper panel, corresponding to a low modulation frequency, reveals a strong broadening of the dip at small magnetic fields when $Q$ is decreased.
The dip becomes deeper at the same time.
In the lower panel, where a high modulation frequency is applied, the shape of the PRCs shows almost no dependence on the pulse strength.
The reason is that the absolute values shown in Fig.~\ref{fig:L_Bext_fm_a} are still much smaller than the spin inertia limit $1/\pi$.
In contrast, the spin inertia signal is in the saturation regime for very low modulation frequencies, 
effectively leading to a decrease of the spin polarization even for relatively large magnetic field. 

\clearpage
\subsection{Increase of spin polarization for detuned pulses}
\label{sec:further_results_detuning}

\begin{figure}[h]
	\centering
	\includegraphics[width=\columnwidth]{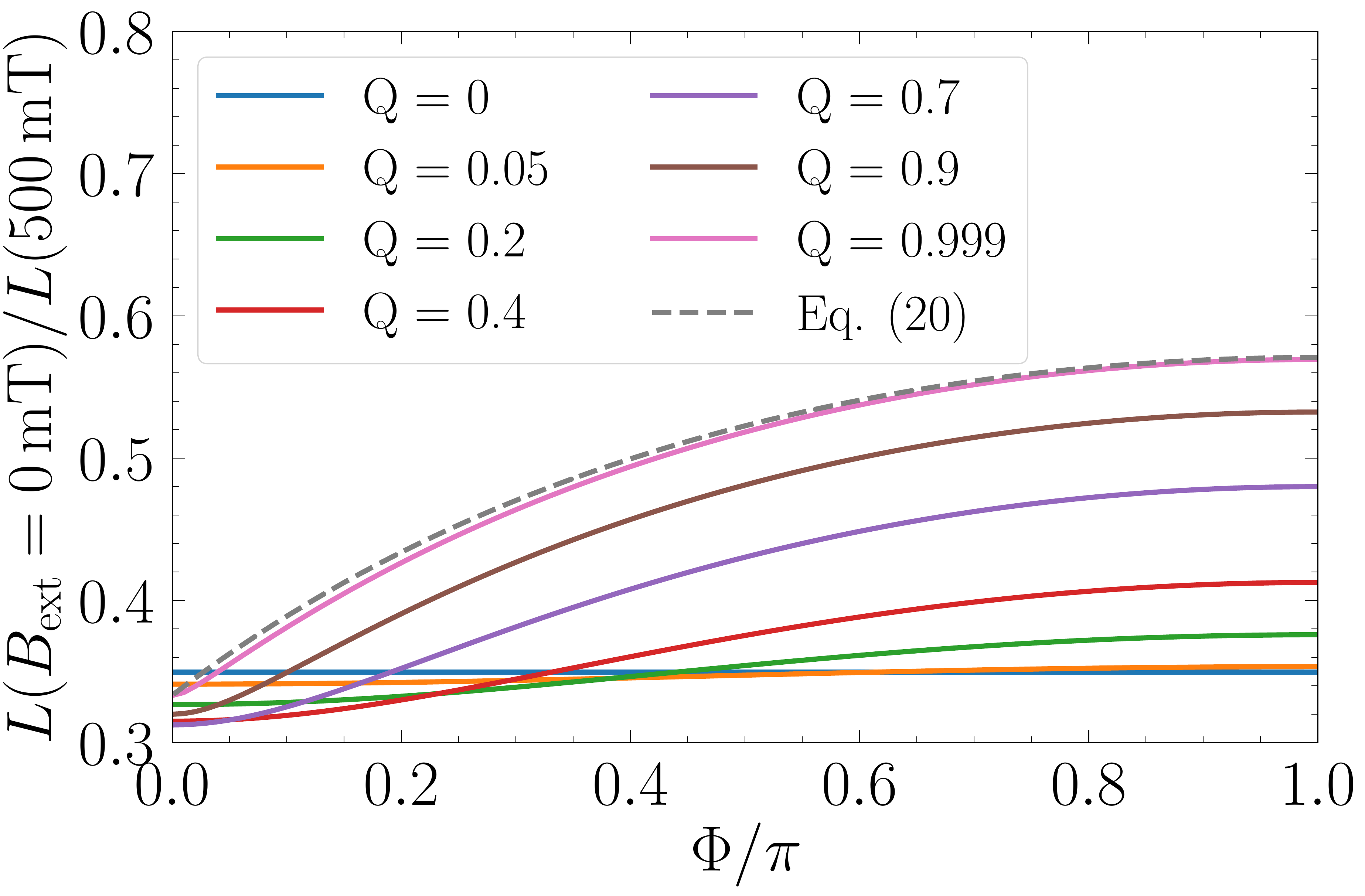}
	\caption{Ratio $L(\Bext = 0\,\mathrm{mT}) / L(500\,\mathrm{mT})$ as a function of $\Phi$ 
		for various pulse strengths $Q$ in the limit of zero modulation frequency $\fm = 0$. 
		The trion spin relaxation time is chosen as $\taust = 0.01$~ns;
		the other parameters are taken from Table~\ref{tab:parameters} for $n$-type QDs.
		The dashed line represents the low-pump-power limit given by Eq.~\eqref{eq:L0_Linfty}.}
	\label{fig:L0_Linfty}
\end{figure}

In order to check the validity of Eq.~\eqref{eq:L0_Linfty}, we perform numerical calculations with the corresponding parameters, see Fig.~\ref{fig:L0_Linfty}.
The convergence to the low pump power limit ($Q \to 1$), which was used for the derivation of Eq.~\eqref{eq:L0_Linfty}, is clearly visible.
In the resonant case $\Phi=0$, we obtain $L(\Bext = 0\,\mathrm{mT})/L(500\,\mathrm{mT})\approx1/3$,
which is the expected value~\cite{merku02,petrov08}. 
Importantly, an increase of $\Phi$ leads to a deviation from the established ratio $1/3$ 
in agreement with Eq.~\eqref{eq:L0_Linfty}. 
At the same time, an increase of the pump power (or a decrease of $Q$) leads to a flattening of this dependence and 
to a slight deviation of $L(\Bext = 0)/L(500\,\mathrm{mT})$ from $1/3$ even at $\Phi=0$. 
The reason is the influence of RSA, which is discussed in detail in Sec.~\ref{sec:RSA}.
As expected for $\pi$ pulses ($Q = 0$), the influence of detuned pulses vanishes completely in accordance with the pulse relation~\eqref{eq:pulse}.
Overall, we find that detuned pulses with $\Phi \ne 0$ lead to an increase of the spin polarization at zero magnetic field when $Q > 0$ is kept constant.

\end{appendix}

\bibliography{inertia.bib}

\end{document}